\documentclass{article}

\usepackage[preprint]{neurips_2026}

\usepackage[utf8]{inputenc} 
\usepackage[T1]{fontenc}    
\usepackage{hyperref}       
\usepackage{url}            
\usepackage{booktabs}       
\usepackage{amsfonts}       
\usepackage{nicefrac}       
\usepackage{microtype}      
\usepackage{xcolor}         
\usepackage{graphicx}       
\usepackage{amsmath}        
\usepackage{algorithm}      
\usepackage{algorithmic}    
\usepackage{caption}        
\usepackage{multirow}       

\title{LPG: Balancing Efficiency and Policy Reasoning in Latent Policy Guardrails}

\author{%
Nanxi Li \qquad Zhengyue Zhao \qquad Chaowei Xiao\\[2pt]
Johns Hopkins University
}

\begin{document}

\maketitle

\begin{abstract}
Guardrails are a critical safety layer for modern AI systems, but their operating regime is changing. As LLMs are deployed as customized assistants, safety policies are increasingly specified at inference time by users, organizations, or regulatory contexts. This makes safety enforcement fundamentally dynamic: the guardrail should adapt to changing safety policies without retraining. Yet this requirement creates a fundamental tension: faithfully judging complex policy contexts demands reasoning capability, while practical deployment requires low-latency responses.
We introduce Latent Policy Guardrail (LPG), a guardrail framework that learns \textbf{semantic latent deliberation} over dynamic policies. LPG compresses the internal deliberation needed for intent interpretation and policy grounding into continuous states supervised by decision-relevant semantics. At inference time, it generates only a compact verdict anchored to the violated policy clauses, preserving auditability while avoiding the latency of explicit reasoning.
Across policy guardrail benchmarks, LPG-4B reaches \textbf{84.5\%} average safety accuracy and \textbf{77.9\%} F1 by compressing deliberation into just \textbf{10} latent tokens, outperforming the strongest dynamic baseline while running roughly \textbf{11\boldmath$\times$} faster than Qwen3-4B-Thinking under the single-sample evaluation setup.
Code and data are available at \url{https://github.com/SaFo-Lab/Latent_Policy_Guard}.
\end{abstract}
\section{Introduction}

Large language models (LLMs) now power customer-facing chatbots~\citep{li2025beyond}, autonomous agents~\citep{wang2024survey}, and healthcare assistants~\citep{rabbani2025generative}, and ensuring that their outputs respect deployment-specific norms has become a central safety concern~\citep{niknazar2024building, bai2022constitutional}. The dominant defense is the \emph{guardrail model}, a lightweight classifier that screens prompts and responses against a safety taxonomy~\citep{inan2023llamaguard, han2024wildguard, zeng2024shieldgemma}. Such models are typically trained on fixed taxonomies (e.g., the OpenAI Moderation API) and have been deployed at scale precisely because they are fast and decisive.

Fixed taxonomies, however, are increasingly out of step with how real systems are deployed. An HR chatbot must prohibit hiring discrimination, a financial advisor must enforce regulatory disclaimers, an educational platform must filter age-inappropriate content. The rise of agentic AI multiplies the demand further, as rules are attached to specific tools, tasks, and tenants~\citep{luo2025agrail, liu2026agentdog}. These constraints cannot all be enumerated at pre-training time, and they evolve faster than any retraining cycle. Recent \emph{policy-aware} guardrails~\citep{hoover2025dynaguard, lin2026xguard} address this by accepting natural-language policies at inference time and judging each interaction against the supplied rules, with no retraining required when the policy changes.

Policy-awareness, however, shifts the burden onto \emph{reasoning}: the guardrail must read a potentially long policy document, identify which clauses apply, and produce a verdict grounded in those clauses, yet existing approaches sit on opposite ends of an unfavorable trade-off. Non-reasoning guardrails such as DynaGuard~\citep{hoover2025dynaguard} are fast but brittle (Section~\ref{sec:investigation}): their verdicts swing by up to $7.4$ accuracy points under simple permutations of the policy list, and a counterfactual probe that drops only the rule an unsafe sample violates fails to flip their verdict, betraying reliance on positional artifacts and content priors rather than the supplied clauses. Reasoning guardrails such as GuardReasoner~\citep{liu2025guardreasoner} and ThinkGuard~\citep{wen2025thinkguard} buy accuracy with explicit chains of thought that, in our measurements, run more than an order of magnitude slower, prohibitive for real-time moderation pipelines. An effective guardrail must therefore anchor each verdict in the specific violated clause and do so independently of surface position, both signatures of \emph{deep} policy understanding, yet the cost of explicit deliberation rules out solutions that rely on long verbal rationales.

A natural remedy is to move the deliberation off the surface tokens. Recent work has shown that LLM reasoning can be performed in continuous latent space rather than via discrete tokens~\citep{hao2024coconut, he2026latentcomputation, amos2026latentreasoning}, yielding speedups on math and planning. Adopting this idea for guardrails, however, is far from a drop-in transfer, and is where the central technical question of this paper lies. In math, latent thoughts are useful because the answer is a deterministic function of an algorithmic working memory, so token-level reconstruction of a teacher's rationale is a natural training target~\citep{hao2024coconut, shen2025codi}. Safety moderation has no comparable algorithmic substrate: the answer is a function of which clause in a heterogeneous, user-supplied policy is implicated, and of the latent intent behind a possibly disguised request, both paraphrastically variable across deployments. Forcing the latents to reproduce specific tokens of a teacher's rationale would over-constrain them, while leaving them unsupervised reduces them to opaque computation buffers untethered from policy content. We argue that what a guardrail actually needs is \textbf{semantic latent deliberation}: each latent stage should retain the decision-relevant gist of its reasoning, the user's underlying intent and the policy clauses it touches, so that a short latent rollout can stand in for a long verbal one without losing policy grounding. This contract, together with the need to anchor verdicts to specific clauses, calls for a guardrail-specific latent design that prior continuous-thought work does not address.

We instantiate semantic latent deliberation in the proposed Latent Policy Guardrail (LPG), a framework that reasons \emph{deeply} but \emph{fast} over user-supplied policies. The two components below are not independent contributions but coupled instruments of one underlying principle, that every reasoning step, whether explicit text or compressed latent, must commit to specific policy semantics rather than diffuse safety priors:

\textbf{Structured reasoning} (Section~\ref{sec:structured}). Rather than unconstrained free-form rationales, we impose a three-stage evaluation template: (1)~\emph{intent analysis}, examining the conversation history to surface jailbreaks hidden behind role-play, hypotheticals, or indirect phrasing; (2)~\emph{policy analysis}, selectively identifying the subset of relevant clauses rather than enumerating all $K$ items, encouraging semantic matching; and (3)~\emph{verdict formulation}, producing the final decision \emph{anchored} to the violated indices~$P^*$. The format simultaneously grounds reasoning in policy content and yields auditable, machine-parseable verdicts.

\textbf{Latent reasoning tailored to policy grounding} (Section~\ref{sec:latent}). Building on continuous-thought reasoning~\citep{hao2024coconut, he2026latentcomputation, amos2026latentreasoning}, we compress the intent and policy stages into continuous hidden states. The harder half of the problem is supervising what those states carry; three coupled design choices, none of which arise in math-style continuous-thought reasoning, realize the semantic contract above. \emph{(i) Stage-aligned latent slots.} Instead of a single undifferentiated continuous-thought buffer, we allocate separate latent budgets to the intent and policy stages, so that the two reasoning sub-tasks live in separable subspaces. \emph{(ii) Semantic-content supervision.} Rather than reconstructing teacher tokens, we train each slot to preserve the gist of its stage by reconstructing a compact teacher \emph{summary}, routed through the same base LM that consumes the latents, which directly shapes the latents to be policy-meaningful representations, not opaque computation buffers. \emph{(iii) Teacher hidden-state distillation at decision-relevant positions}, applied at both stage boundaries and the verdict-onset position, so that the latent pathway converges on the same internal computation that produces the teacher's explicit decision. The full multi-objective loss is detailed in Section~\ref{sec:training}.

Empirically, semantic latent deliberation delivers a substantially better accuracy--latency Pareto frontier (Sections~\ref{sec:results} and \ref{sec:ablation}). On in-distribution benchmarks, LPG-4B reaches $84.5\%$ average safety accuracy and $77.9\%$ F1, outperforming the strongest dynamic baseline and reasoning baselines while running $\sim$$11\times$ faster than Qwen3-4B-Thinking. The gains transfer out of distribution: LPG reaches $96.4\%$ F1 on HarmBench and remains the best dynamic guardrail on WildGuardTest. Ablations isolate the answer-position distillation sub-loss as the load-bearing supervision signal (removing it costs nearly $20$ accuracy points), and a latent-budget sweep shows that just $10$ latent tokens already approach full explicit reasoning at $\sim$$5\times$ lower latency, with diminishing returns beyond $20$ tokens. To our knowledge, LPG is the first framework to apply latent reasoning specifically to policy-grounded safety moderation, demonstrating that latent compression is a practical bridge between fast static classifiers and slow but expressive policy-reasoning guardrails.

\section{Related Work}
\label{sec:related}

\paragraph{LLM guardrail models.}
Llama Guard~\citep{inan2023llamaguard} established fine-tuning LLMs on fixed safety taxonomies. Subsequent work scaled this across model families (ShieldGemma~\citep{zeng2024shieldgemma}, Qwen3Guard~\citep{zhao2025qwen3guard}) and model sizes (Llama Guard 3~\citep{fedorov2024llamaguard1b}, PolyGuard~\citep{kumar2025polyguard}), yet all require retraining for novel policies. Reasoning-based methods like GuardReasoner~\citep{liu2025guardreasoner} and ThinkGuard~\citep{wen2025thinkguard} improve accuracy via explicit chain-of-thought~\citep{li2024safetyanalyst,kang2024r}, but incur large latency overhead. LPG addresses this trade-off by replacing verbose textual reasoning with compact latent representations.

\paragraph{Dynamic policy-aware guardrails.}
A growing line of work conditions safety judgments on user-specified policies provided at inference time. DynaGuard~\citep{hoover2025dynaguard} introduces a framework where natural-language policies are included in the prompt, along with a companion benchmark, DynaBench, for evaluation. YuFeng-XGuard~\citep{lin2026xguard} proposes a hierarchical inference strategy that decouples policy from risk perception, enabling policy updates without retraining. In the agent safety domain, AGrail~\citep{luo2025agrail} and AgentDoG~\citep{liu2026agentdog} develop guardrails that monitor agent trajectories against task-specific safety constraints. Our work shares the policy-aware philosophy of these methods but focuses on the largely unexplored question of how to \emph{efficiently} reason over policies, rather than relying on computationally expensive explicit generation.

\paragraph{Latent reasoning in language models.}
Recent work has demonstrated that LLMs can perform effective reasoning in continuous latent space rather than through discrete token generation. COCONUT~\citep{hao2024coconut} proposes a paradigm where continuous thoughts encode multiple alternative reasoning paths. CODI~\citep{shen2025codi} shows that aligning the student's hidden state at a single token to the teacher's answer-position token, recovers explicit-CoT accuracy on math. Follow-up studies show that steering a single latent reasoning feature can improve accuracy without explicit CoT~\citep{he2026latentcomputation}, that supervised thinking states in embedding space achieve competitive performance~\citep{amos2026latentreasoning}, and that latent reasoning transfers across domains~\citep{ye2026latentchem}. Emergent search behaviors have also beaen observed in latent reasoning models~\citep{cui2026emergentsearch}. AISA~\citep{song2026aisa} leverages latent safety awareness through attention-head analysis for jailbreak defense without fine-tuning. Concurrent work DRAFT~\citep{wang2026draft} compresses agent-trajectory safety into a single continuous latent draft supervised end-to-end by binary cross-entropy on static policies. Our LPG instead targets \emph{user-supplied} policies at inference time, which motivates our stage-aligned latent slots, semantic-content supervision via teacher distillation and summary reconstruction, and clause-anchored verdicts, none of which are addressed by DRAFT.

\section{Investigation Experiments}
\label{sec:investigation}

To validate the importance of policy-aware reasoning in guardrail models, we conduct two preliminary investigation experiments.

\subsection{Performance with and without Policy}

We evaluate Qwen3-4B, Qwen3-4B (Thinking), and DynaGuard-8B under three conditions: \textbf{Full Policy} (standard); \textbf{Remove All Policies}, a degenerate baseline that drops the entire policy list; and the counterfactual \textbf{Remove Violated only}, which drops \emph{only} the policy item(s) an unsafe sample violates and keeps the rest, flipping the ground truth to \texttt{safe} so a policy-grounded model should follow.

\begin{figure}[t]
\centering
\begin{minipage}[t]{0.53\textwidth}
\vspace{0pt}
\centering
\includegraphics[width=\linewidth]{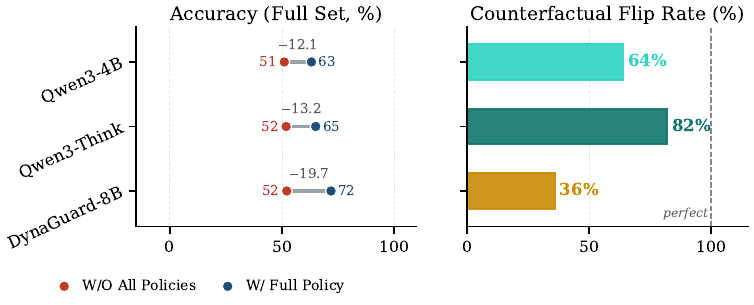}
\captionof{figure}{\textbf{Policy grounding probes.} \textbf{Left:} Accuracy on the full set with vs.\ without the policy. \textbf{Right:} counterfactual flip rate on the unsafe subset: fraction of samples whose verdict correctly flips to safe after only the violated rule(s) are removed; higher = the model anchors on the specific violated clause.}
\label{fig:policy_comparison}
\end{minipage}\hfill
\begin{minipage}[t]{0.45\textwidth}
\vspace{0pt}
\centering
\footnotesize
\setlength{\tabcolsep}{4pt}
\renewcommand{\arraystretch}{1.4}
\begin{tabular}{l c c c}
\toprule
\textbf{Model} & \multicolumn{2}{c}{\textbf{Acc (\%)}} & \textbf{CR (\%)} \\
\cmidrule(lr){2-3}
& \textbf{Orig.} & \textbf{Shuf.} & \textbf{mean$\pm$std} \\
\midrule
Qwen3-4B          & 75.60 & 72.83 & 79.64$\pm$40.27 \\
Qwen3-4B (Think)  & 77.25 & 77.68 & 89.13$\pm$31.13 \\
GuardReason-3B    & 60.70 & 67.07 & 65.66$\pm$47.49 \\
DynaGuard-8B      & 81.65 & 74.23 & 86.73$\pm$33.93 \\
\bottomrule
\end{tabular}
\captionof{table}{Consistency under policy shuffling on GuardSet-X. \textbf{Orig.}/\textbf{Shuf.}: Safety Accuracy with original / mean of three perturbed orderings; \textbf{CR}: per-sample verdict consistency rate.}
\label{tab:policy_shuffling}
\end{minipage}
\end{figure}

As depicted in Figure~\ref{fig:policy_comparison} left, withholding the full policy drops Accuracy by $12$--$20$ points and collapses F1 from $26.9$/$55.1$/$56.6\%$ to $<$$6\%$ as models default to safe. The counterfactual is more revealing (Figure~\ref{fig:policy_comparison}, right): Qwen3-4B's flip-to-safe rate is $64\%$, Thinking strengthens it to $82\%$, but specialised DynaGuard-8B \emph{collapses} to $36\%$, below its own Full-Policy verdict correctness.

\subsection{Policy Shuffling Experiments}

We test whether models reason over policy content by randomly permuting the policy list under three orderings (original, reversed, random) and measuring per-sample verdict consistency rate (CR) and the Accuracy gap between original and shuffled inputs.

As shown in Table~\ref{tab:policy_shuffling}, mean CR ranges from $65.7\%$ (GuardReasoner-3B) to $89.1\%$ (Qwen3-4B-Think), but the high standard deviations ($31.1$--$47.5$) reveal that many verdicts flip arbitrarily under shuffling, and Accuracy drops by up to $7.4$ points. Adding explicit reasoning does some help: Qwen3-4B-Thinking's CR \emph{increases} from $79.6\%$ to  $89.1\%$, suggesting deliberate thinking can help mitigate it.

\paragraph{Findings and implications.} Both probes diagnose the same failure: existing baselines do not condition on the \emph{specific} active rule set: removing the violated rule does not reliably flip the verdict, and shuffling the rule list does. The regression is sharpest on DynaGuard-8B, the only specialised guardrail in both probes: it is worst on the counterfactual and shows the largest Accuracy drop under shuffling ($-7.4$ pt), indicating that policy-grounded fine-tuning has \emph{eroded} the policy-reasoning ability that instruction-following baselines retain. This motivates the two design choices in Section~\ref{sec:method}: (i) \emph{policy anchoring}, which forces each unsafe verdict to commit to specific violated indices~$P^*$ rather than emit a verdict in isolation, and (ii) a training corpus that uses various policy lists with augumentations, so the model cannot collapse onto positional or memorisation shortcuts.

\section{Method}
\label{sec:method}


\subsection{Problem Formulation}
\label{sec:problem}

Given content to moderate $x$ and a policy document $\mathcal{P} = \{p_1, \ldots, p_K\}$ of $K$ natural-language policy items, the guardrail $f_\theta$ must produce a binary verdict $y \in \{\texttt{safe}, \texttt{unsafe}\}$ and, when $y = \texttt{unsafe}$, the set $P^* \subseteq \{1, \ldots, K\}$ of violated policy indices ($P^* = \emptyset$ when safe). We write the full context as $\mathbf{c} = (\mathcal{P}, x)$. The challenge is that faithfully evaluating $\mathbf{c}$ against all $K$ items demands substantial reasoning, yet must respect a strict latency budget.

\subsection{Structured Evaluation Template}
\label{sec:structured}

Two design questions motivate this template. \emph{First, why constrain reasoning at all?} Free-form chain-of-thought is the obvious baseline~\citep{liu2025guardreasoner, wen2025thinkguard}, but it leaks two pathologies into a guardrail: (i) it generates many tokens that are irrelevant for the verdict, inflating latency, and (ii) it tends to skim policy text and fall back on prior-trained safety priors, which Section~\ref{sec:investigation} shows are insufficient for user-defined rules. \emph{Second, why three stages and in this order?} Many real-world unsafe interactions disguise malicious intent under benign surface forms (role-play, hypotheticals)~\citep{zhao2025armor}. Forcing the model to commit to an \emph{intent} hypothesis before it touches the policy decouples ``what is the user really asking for'' from ``which rule covers it,'' which we find more robust than mixing the two; the final verdict, anchored to specific indices~$P^*$, then forces any positive judgment to point at concrete clauses. Inspired by structured safety alignment~\citep{li2024safetyanalyst, li2025prism}, we therefore constrain the model to a fixed three-stage reasoning trace.

\paragraph{Stage 1: Intent analysis.} The model first analyzes the content to moderate $x$ to uncover the underlying intent, with particular attention to jailbreaks that hide malicious instructions through role-play, hypothetical scenarios, or indirect phrasing.

\paragraph{Stage 2: Policy analysis.} Rather than enumerating all $K$ items, the model selectively identifies a subset $\mathcal{K}_{\text{relevant}} \subset \{1, \ldots, K\}$ of relevant policies and assesses, for each, whether the conversation constitutes a violation. This selective treatment is more efficient than full enumeration and encourages semantic understanding rather than positional pattern matching.

\paragraph{Stage 3: Verdict formulation.} The model emits a compact natural-language string (one of \texttt{``safe''}, \texttt{``unsafe, policy $n$''}, or \texttt{``unsafe, policy $n_1, n_2, \ldots$''}) that combines the binary verdict with $P^*$. The format is parseable by a deterministic regex and significantly shorter than a JSON-formatted alternative, while still anchoring each verdict to specific clauses for downstream auditing.

\subsection{Latent Reasoning via Continuous-Thought Compression}
\label{sec:latent}

While the structured template constrains the format of reasoning, the token count for Stages 1--2 still scales with conversation length and policy count. To reduce inference cost further, we compress these two stages into continuous latent representations, building on the observation that language-space verbalization is often unnecessarily verbose for the underlying computation~\citep{hao2024coconut, shen2025codi}. Stage 3 remains explicit text, preserving interpretability of the final decision.

\paragraph{Why guardrail latents need a different supervision signal.}
Prior latent-reasoning works on math and planning~\citep{hao2024coconut, shen2025codi, wang2025system15} treat the latent thought as a compressed surrogate for the algorithm's intermediate computations: token-level reconstruction (or its information-theoretic equivalent) is a sensible target because the answer is a deterministic function of the working memory. Guardrail reasoning is qualitatively different. The verdict for an unsafe interaction depends on (a)~which natural-language clause is violated and (b)~the latent intent behind a possibly disguised request, both of which are paraphrastically variable across the policies a deployed model will encounter. Forcing the latents to reproduce specific tokens of the teacher's rationale would over-constrain them; what we actually need is for each latent slot to retain the decision-relevant semantic content of its stage. We translate this view into three concrete design choices: (i) two \emph{stage-aligned} latent slots that mirror the structured template, so that intent and policy-violation reasoning live in separable subspaces rather than sharing one undifferentiated buffer; (ii) \emph{semantic-content supervision} via teacher-summary reconstruction routed through the same base LM that consumes the latents, shaping the latents so that the LM's continuation matches the gist of the stage; and (iii) \emph{teacher hidden-state distillation} at stage boundaries and at the verdict-onset position, which directly transfers the teacher's decision-producing representations into the student's latent pathway. Both signals are formalized in Section~\ref{sec:training}. The combination yields short latents that nonetheless remain aligned with policy content rather than acting as opaque computation buffers.

\begin{figure}[t]
\centering
\includegraphics[width=\linewidth]{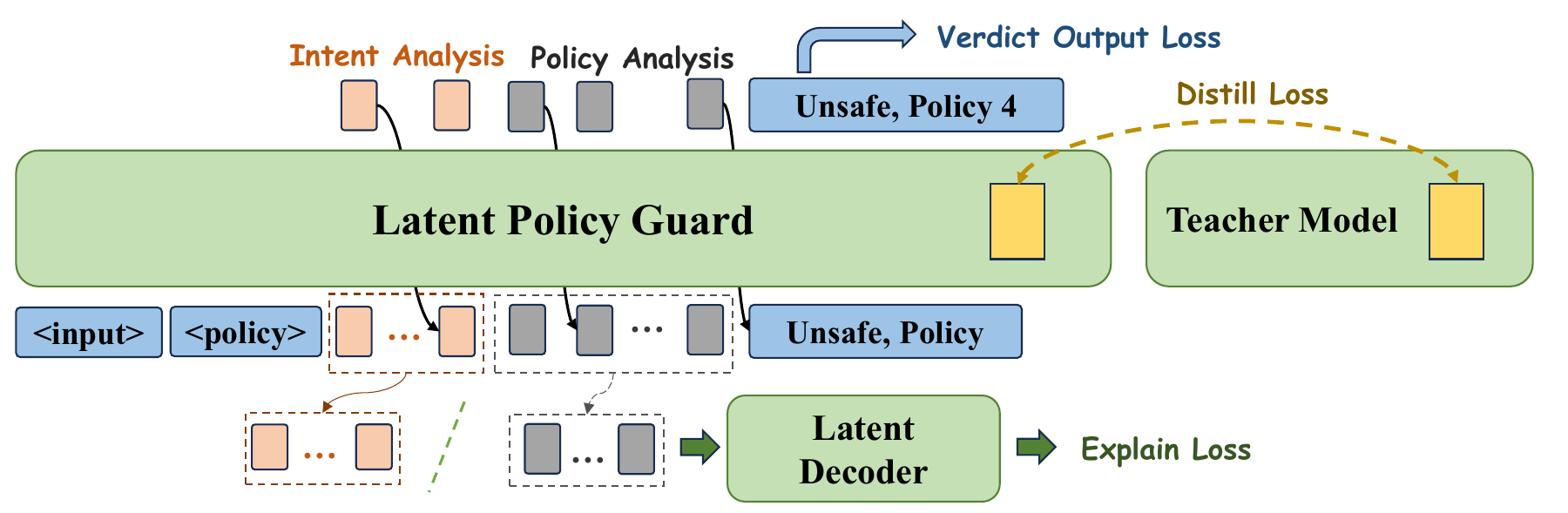}
\caption{Overview of LPG. The student compresses intent analysis and policy analysis into two latent stages, then decodes the verdict and violated policy indices.}
\label{fig:overview}
\end{figure}

\paragraph{Latent token replacement.}
Let $\mathbf{h}_t^{(L)} \in \mathbb{R}^d$ denote the top-layer hidden state at position $t$. We replace explicit token generation for Stages 1--2 with \emph{latent tokens} whose embeddings are set directly from the previous hidden state via a learned projection: $\mathbf{e}_{t+1} = \text{Proj}(\mathbf{h}_t^{(L)})$, yielding a continuous thought chain that proceeds in latent space without materializing into discrete tokens. The supervision signals that shape these latents (a stage-summary reconstruction loss and teacher hidden-state distillation) are introduced together in Section~\ref{sec:training}.

\subsection{Policy Anchoring}
\label{sec:anchoring}

Requiring the model to emit the violated indices $P^*$ acts as both an inductive bias and a training signal. It prevents vague safety judgments untied to any rule in $\mathcal{P}$, enables automated auditing of which clauses each flagged item cites, and provides fine-grained supervision beyond the binary verdict, encouraging genuine policy comprehension rather than surface pattern matching.

\subsection{Training Corpus}
\label{sec:corpus}

LPG is trained on a $40k$-record mixture combining two policy-grounded datasets (DynaBench~\citep{hoover2025dynaguard} and GuardSet-X~\citep{wen2025towards}) with a diverse set of public safety moderation datasets, including BeaverTails~\citep{ji2023beavertails}, Aegis-AI Content Safety v2~\citep{ghosh2025aegis}, SaladBench~\citep{li2024saladbench}, Toxic-Chat~\citep{lin2023toxicchat}, and XSTest v2~\citep{rottger2023xstest}. To unify these heterogeneous sources, we reorganize the GuardSet-X from single policy into a shared multi-policy book format, and for the general safety moderation datasets we write policy items from their native safety taxonomies. All sources are then converted into the single $(\mathcal{P}, x, y, P^*)$ schema with teacher-generated structured reasoning (Section~\ref{sec:structured}); the final mixture is $45.5\%$ safe / $54.5\%$ unsafe with mean $0.63$ violations per record. The GuardSet-X test split is held out for evaluation.

The corpus is built around a single design principle: no training example should see the same policy list twice, so the latent pathway cannot collapse onto policy ordering or memorized rule strings. To enforce this, (i)~each source's native taxonomy is rewritten into a single-sentence ``policy book''; (ii)~every record is paired with a policy list, always including the violated rule(s) when unsafe and (iii)~all rationales are produced by a Qwen3-32B teacher~\citep{yang2025qwen3} conditioned on the ground-truth verdict $(y, P^*)$, which eliminates reasoning--label drift before it can propagate into the student's latents. The full curation pipeline is shown in Appendix~\ref{sec:training_details}.

\subsection{Training Procedure}
\label{sec:training}

We supervise a fast student that reasons in latent space using a trained teacher $f_{\theta_T}$ that produces explicit structured reasoning. Training is single-phase end-to-end with four loss terms; full pseudocode is in Appendix~\ref{sec:training_algorithm}.

\paragraph{Verdict output loss.}
The student emits the compact verdict string after the latent rollout, preceded by a learned \texttt{<eot>} marker. We apply standard next-token cross-entropy on the verdict positions:
\begin{equation}
\mathcal{L}_{\text{out}} = -\sum_{t \in T_{\text{verdict}}} \log P_\theta(r_t \mid r_{<t}, \mathbf{c}).
\label{eq:out_loss}
\end{equation}

\paragraph{Teacher hidden-state distillation.}
We distill the teacher's reasoning trajectory at two complementary positions: the \emph{answer position}, where we align all $L$ layers between student and teacher at the verdict-onset token (transferring the verdict-producing computational state across the residual stack), and the \emph{stage boundaries} \texttt{</Intent>}, \texttt{</Risk>}, where we align each stage's final top-layer latent to the teacher's hidden state at the matching boundary token. The answer-position term is closely modeled after CODI's designated-token self-distillation~\citep{shen2025codi}, which shows that aligning student and teacher hidden states at a single answer-position token suffices to inherit explicit-CoT accuracy on math; we extend that recipe to all $L$ layers and complement it with the stage-boundary term so that each stage-aligned latent slot is anchored to the teacher's representation at its corresponding boundary token, which CODI's single-point design does not provide. Both alignments use SmoothL1 normalized by the teacher's per-vector standard deviation $\sigma(\cdot)$ for layer-scale invariance:
\begin{equation}
\mathcal{L}_{\text{distill}}^{\text{ans}} = \frac{1}{L}\sum_{l=1}^{L} \frac{\text{SmoothL1}\!\left(\mathbf{h}^{(l)}_{\text{stu}}[t_a],\, \mathbf{h}^{(l)}_{\text{tea}}[t_a']\right)}{\sigma\!\left(\mathbf{h}^{(l)}_{\text{tea}}[t_a']\right)},
\quad
\mathcal{L}_{\text{distill}}^{\text{stage}} = \sum_{k \in \{1,2\}} \frac{\text{SmoothL1}\!\left(\mathbf{z}^{(k)}_{m_k},\, \mathbf{h}^{(L)}_{\text{tea}}[b_k]\right)}{\sigma\!\left(\mathbf{h}^{(L)}_{\text{tea}}[b_k]\right)},
\label{eq:distill}
\end{equation}
where $t_a, t_a'$ are answer positions in the student/teacher sequences, $\mathbf{z}^{(k)}_{m_k}$ is the final latent of stage $k$, and $b_k$ is the closing-tag position for stage $k$ in the teacher. The two sub-losses target complementary cross-sections of the residual stream (all-layer alignment at one point in time vs.\ top-layer alignment at two stage transitions), and combine as $\mathcal{L}_{\text{distill}} = \mathcal{L}_{\text{distill}}^{\text{ans}} + \beta\,\mathcal{L}_{\text{distill}}^{\text{stage}}$ with $\beta{=}0.1$ (Appendix~\ref{sec:training_details}).

\paragraph{Summary reconstruction for semantic supervision.}
Distillation transfers the teacher's representations point-wise, but it does not by itself ensure that each latent slot remains semantically interpretable to the base LM. To preserve that grounding, we add a stage-summary reconstruction objective in place of the token-level reconstruction loss used by math-style latent reasoning. Let $\mathbf{Z}^{(k)} \in \mathbb{R}^{m_k \times d}$ denote the $m_k$ latent tokens of stage $k \in \{1,2\}$. The teacher provides an essential-information target $\mathbf{s}^{(k)}$: the detected intent (including any hidden malicious goal) for $k{=}1$, and the relevant policy clauses plus a concise violation rationale for $k{=}2$. We project $\mathbf{Z}^{(k)}$ through a learned multi-layer projector $\text{Proj}_k$ (Linear--GELU--LayerNorm--Dropout), prepend the result to the teacher-forced summary embeddings, and reconstruct $\mathbf{s}^{(k)}$ via next-token prediction through the base LM:
\begin{equation}
\mathcal{L}_{\text{explain}} = \sum_{k \in \{1,2\}} \left(-\sum_{t} \log P_{\theta}\!\left(s^{(k)}_t \mid f_\theta\!\left([\text{Proj}_k(\mathbf{Z}^{(k)}); \mathbf{s}^{(k)}_{<t}]\right)\right)\right).
\label{eq:explain_loss}
\end{equation}
Routing reconstruction through the base LM rather than a separate decoder directly shapes the model's internal representations of the latent tokens. The projectors $\{\text{Proj}_k\}$ are used only during training; at inference, the model operates purely in latent space and decodes the verdict directly.

\paragraph{Reference reasoning loss.}
A separate forward pass through the teacher's full explicit reasoning sequence (\texttt{<Intent>}, \texttt{<Risk>}, \texttt{<Output>}) supplies a cross-entropy backbone-LM regularizer that prevents catastrophic forgetting of explicit reasoning capability:
\begin{equation}
\mathcal{L}_{\text{ref}} = -\sum_{t \in T_{\text{reasoning}}} \log P_\theta(r_t \mid r_{<t}, \mathbf{c}).
\label{eq:ref_ce}
\end{equation}

\paragraph{Joint objective.}
The four losses combine in a single phase:
\begin{equation}
\mathcal{L} = \lambda_{\text{out}} \mathcal{L}_{\text{out}} + \lambda_{\text{distill}} \mathcal{L}_{\text{distill}} + \lambda_{\text{ref}} \mathcal{L}_{\text{ref}} + \lambda_{\text{explain}} \mathcal{L}_{\text{explain}},
\label{eq:total_loss}
\end{equation}
with $\mathcal{L}_{\text{explain}}$ from Eq.~\ref{eq:explain_loss}. The explain projectors are discarded at inference; all other components remain active.

\section{Main Results}
\label{sec:results}

\subsection{Policy Safeguarding Evaluation}
\label{sec:indist}

We first compare LPG against existing guardrails on the two policy-grounded benchmarks that match the training distribution: GuardSet-X and the challenging augmented DynaBench split. DynaBench is itself a demanding policy-grounded benchmark, with each example carrying on average $13.8$ policy items and up to $91$ in the heaviest configurations, which requires the model to identify the relevant policies from a large set of distractors. Inspired by the failure modes surfaced in Section~\ref{sec:investigation}, we further augment DynaBench along two axes: (i)~each policy list is randomly shuffled, and (ii)~additionally include counterfactual variants in which the violated rule is removed from the policy list (so the corresponding label flips to safe). Details for this augumentation are in Appendix~\ref{sec:datasets_details}. The shuffling tests whether the model anchors on policy content rather than positional cues, and the counterfactuals test whether the model's verdicts are actually grounded in the specific violated clauses rather than relying on prior safety priors.

\begin{table}[t]
\small
\setlength{\tabcolsep}{4pt}
\caption{Results on GuardSet-X and DynaBench. \textbf{Acc}: Safety Accuracy (binary verdict correctness). \textbf{F1}: Safety F1 (violation detection). \textbf{Lat.}: average inference time per sample in milliseconds (single-sample batch on an A100-80GB; see Section~\ref{sec:experimental_setup}). \textbf{Avg.}: per-metric mean across the two benchmarks.}

\label{tab:results}
\begin{center}
\begin{tabular}{lccc|ccc|ccc}
\toprule
\textbf{Model} & \multicolumn{3}{c|}{\textbf{GuardSet-X}} & \multicolumn{3}{c|}{\textbf{DynaBench (Aug)}} & \multicolumn{3}{c}{\textbf{Avg.}} \\
\cmidrule(lr){2-4} \cmidrule(lr){5-7} \cmidrule(lr){8-10}
& \textbf{Acc} & \textbf{F1} & \textbf{Lat.} & \textbf{Acc} & \textbf{F1} & \textbf{Lat.} & \textbf{Acc} & \textbf{F1} & \textbf{Lat.} \\
\midrule
Qwen3-4B             & 75.60 & 71.86 & 249  & 63.05 & 26.88 & 227  & 69.33 & 49.37 & 238 \\
Qwen3-4B-Thinking    & 77.25 & 74.07 & 9747 & 65.01 & 55.12 & 7019 & 71.13 & 64.60 & 8383 \\
DynaGuard-4B         & 81.00 & 79.33 & 222  & 59.48 & 53.19 & 242  & 70.24 & 66.26 & 232 \\
DynaGuard-8B         & 81.65 & 79.73 & 1501 & 71.82 & 56.59 & 1091 & 76.74 & 68.16 & 1296 \\
GuardReasoner-3B     & 60.70 & 38.30 & 3537 & 58.74 & 31.28 & 3993 & 59.72 & 34.79 & 3765 \\
GuardReasoner-8B     & 63.16 & 22.48 & 4072 & 57.66 & 27.77 & 3969 & 60.41 & 25.13 & 4021 \\
\textbf{LPG-4B (Ours)} & \textbf{96.85} & \textbf{96.88} & 625 & \textbf{72.19} & \textbf{58.82} & 871 & \textbf{84.52} & \textbf{77.85} & 748 \\
\bottomrule
\end{tabular}
\end{center}
\end{table}

Table~\ref{tab:results} shows that LPG delivers the strongest performance across both policy-grounded benchmarks, reaching 84.52/77.85 average Acc/F1. For comparison on DynaBench (Aug), which is in-distribution for both LPG and DynaGuard and tests robustness under shuffled and counterfactual policy variants: LPG ranks first at 72.19/58.82. The gain is narrow in absolute terms but the relative pattern (DynaGuard-4B drops more Acc compared with the original DynaBench in Appendix Table~\ref{tab:results_dynabench_old}, while LPG retains its lead) suggests LPG's policy anchoring is more robust to surface-form perturbation. These gains come with a favorable accuracy--efficiency trade-off: at 748~ms average latency, LPG is roughly $11\times$ faster than the explicit-reasoning Qwen3-4B-Thinking and $5.4\times$ faster than GuardReasoner-8B under the same single-sample setup, narrowing the latency gap to lightweight classifiers while preserving the benefits of policy-grounded reasoning. A dedicated decoding-variance sweep in Appendix~\ref{sec:variance} (Table~\ref{tab:variance}) confirms these gains are statistically stable: across $n{=}15$ runs per (model, dataset) cell, the $95\%$ CI half-widths are at most $1.34$ points, well below the cross-model gaps reported here.

\subsection{Out-of-Distribution Evaluation}
\label{sec:ood}

To assess whether LPG generalizes beyond its training distribution, we evaluate on three benchmarks that were not seen during training: two general safety datasets, \textbf{HarmBench}~\citep{mazeika2024harmbench} and \textbf{WildGuardTest}~\citep{han2024wildguard}, and one single policy-grounded benchmark, \textbf{PolicyGuardBench}~\citep{wen2025towards}. HarmBench and WildGuardTest ship without explicit policy strings, so we synthesize per-example policy lists from each benchmark's native taxonomy; PolicyGuardBench supplies $\sim$60k policy-violation labels over web-agent trajectories. We compare LPG against a broad cross-section of baselines spanning three families: instruction-following models (Qwen3-4B~\citep{yang2025qwen3}, GPT-4o~\citep{openai2024gpt4o}), static policy guardrails (ShieldGemma~\citep{zeng2024shieldgemma}, GuardReasoner~\citep{liu2025guardreasoner}), and dynamic policy guardrails (DynaGuard~\citep{hoover2025dynaguard}). 

\begin{table}[t]
\small
\caption{Out-of-distribution Safety F1 (\%) on HarmBench, WildGuardTest, and PolicyGuardBench. Entries marked with $^{\dagger}$ indicate that the corresponding benchmark is in-distribution for that model.}
\label{tab:ood_results}
\begin{center}
\begin{tabular}{llcccc}
\toprule
\textbf{Family} & \textbf{Model} & \textbf{HarmBench} & \textbf{WildGuardTest} & \textbf{PolicyGuardBench} \\
\midrule
\multirow{2}{*}{Instruction-following}
& Qwen3-4B            & 83.21 & 78.88 & 53.48  \\
& GPT-4o              & 82.27 & 80.87 & 87.77 \\
\midrule
\multirow{3}{*}{Static guardrail}
& LlamaGuard3-8B      & 67.96 & 68.47 & 59.52  \\
& ShieldGemma-9B      & 67.96 & 57.74 & 34.72  \\
& GuardReasoner-8B    & 91.86 & $89.17^{\dagger}$ & 77.64 \\
\midrule
\multirow{2}{*}{Dynamic guardrail}
& DynaGuard-4B        & 86.32 & 81.09 & 77.02 \\
& \textbf{LPG-4B (Ours)} & 96.44 & 84.09 & 77.85 \\
\bottomrule
\end{tabular}
\end{center}
\end{table}

Table~\ref{tab:ood_results} shows that LPG transfers beyond its training distribution, scoring 96.44 on HarmBench and 84.09 on WildGuardTest, the strongest result among models for which the benchmark is out-of-distribution. On the policy-grounded PolicyGuardBench, LPG reaches 77.85 F1, edging DynaGuard-4B (77.02) and trailing only the much larger GPT-4o, indicating that the benefits of latent policy reasoning are not confined to the training setup. The comparison also highlights a clear pattern: methods tied to fixed taxonomies degrade sharply when the policy space changes, with ShieldGemma-9B falling to 34.72 and LlamaGuard3-8B to 59.52 on PolicyGuardBench, whereas LPG remains robust because it reasons directly over user-supplied rules. Overall, these results suggest that LPG generalizes across both conventional safety moderation and policy-grounded agent settings while retaining the efficiency advantages of latent reasoning.

\section{Ablation Study}
\label{sec:ablation}


\begin{figure}[t]
\centering
\includegraphics[width=\linewidth]{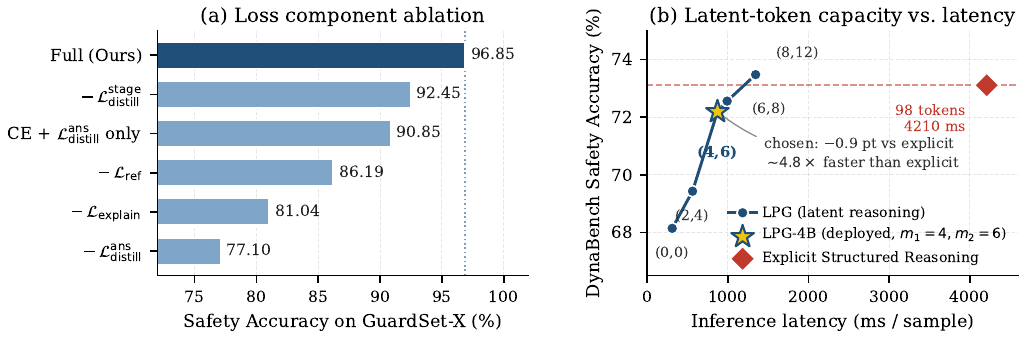}
\caption{\textbf{Ablation studies.} \textbf{(a)} Loss-component ablation: GuardSet-X Safety Accuracy when each term (or distillation sub-loss) is removed from Eq.~\ref{eq:total_loss}; dotted line marks the full model. \textbf{(b)} Latent-token capacity sweep on DynaBench; the red diamond is the explicit-reasoning baseline.}
\label{fig:ablation}
\end{figure}

\paragraph{Loss components.} The verdict output loss ($\mathcal{L}_{\text{out}}$) and the answer-position sub-loss ($\mathcal{L}_{\text{distill}}^{\text{ans}}$) are the two indispensable components: the former teaches the output format and the latter transfers the teacher's verdict-producing hidden states to the student's latent pathway. Removing $\mathcal{L}_{\text{distill}}^{\text{ans}}$ drops accuracy by nearly $20$ points, the largest single regression we observe. The remaining components provide complementary semantic supervision: $\mathcal{L}_{\text{explain}}$ encourages each latent stage to retain the information for its reasoning function, the stage-boundary sub-loss $\mathcal{L}_{\text{distill}}^{\text{stage}}$ supplies a direct hidden-state target for each stage's final latent token, and $\mathcal{L}_{\text{ref}}$ regularizes the backbone LM and prevents catastrophic forgetting of the explicit reasoning capability that underpins the latent representations.

\paragraph{Latent token count.} We vary the latent budget $(m_1, m_2)$ allocated to the intent and risk stages, keeping their ratio at approximately $2{:}3$ to match the explicit reasoning length proportion. The $(0,0)$ endpoint serves as a useful reference: with no latent positions, the model collapses to a no-thinking baseline, i.e., a DynaGuard-style direct-verdict guardrail fit to the LPG corpus. It scores $68.14\%$ on DynaBench, roughly $4$ points below the deployed $(4,6)$ setting, isolating the gain attributable to the latent reasoning pathway from that of the training data alone. Performance rises with the latent budget, where LPG reaches $73.48\%$ on DynaBench, marginally exceeding the explicit-reasoning baseline while running at $1344$\,ms per sample. Each additional latent token, however, triggers one extra forward pass through the base LM, so the latency curve grows linearly, and the explain loss becomes harder to optimize at larger budgets. We therefore deploy the configuration that lands within roughly one point of the saturated setting: $(m_1, m_2) = (4, 6)$ achieves $72.19\%$ at $871$\,ms per sample ($0.92$ points below the explicit baseline at ${\sim}4.8{\times}$ lower latency), which we judge the best accuracy-latency trade-off for production deployment.

\section{Conclusion}
\label{sec:conclusion}

We presented \textbf{LPG}, a policy-grounded guardrail that compresses two-stage safety reasoning (intent analysis and risk assessment against user-supplied policies) into a small budget of continuous latent tokens, supervised by a teacher-distilled multi-objective loss. Across policy guardrail benchmarks, LPG matches or exceeds the strongest dynamic and reasoning-based baselines while running roughly $5\times$ faster than the strongest reasoning baseline and $11\times$ faster than Qwen3-4B-Thinking on a single-sample setup. We hope this work motivates further study of latent reasoning as a practical bridge between fast static classifiers and slow but expressive reasoning guardrails, particularly in deployment regimes where policies are dynamic and latency budgets are tight.

\clearpage
\bibliographystyle{plain}
\bibliography{references}

\clearpage
\appendix

\section*{Supplementary Material}

\section{Experimental Setup}
\label{sec:experimental_setup}

\subsection{Baseline Models}

We evaluate the following baseline models:

\textbf{Static Policy Baselines.}
Llama Guard 3~\citep{fedorov2024llamaguard1b} and ShieldGemma-2B~\citep{zeng2024shieldgemma} are static guardrail models trained on fixed safety taxonomies. They cannot adapt to custom policies.

\textbf{Dynamic Policy-Aware Baselines.}
DynaGuard-4B/8B~\citep{hoover2025dynaguard} introduces a framework for user-defined safety policies. Qwen3-4B~\citep{yang2025qwen3} and GPT-4o~\citep{openai2024gpt4o} are evaluated zero-shot with policies provided in-context. Qwen3-4B-Thinking extends Qwen3 with explicit thinking tokens for reasoning.

\textbf{Reasoning-Based Baselines.}
GuardReasoner-3B/8B~\citep{liu2025guardreasoner} trains on 127K samples with explicit chain-of-thought rationales and applies hard-sample DPO. ThinkGuard~\citep{wen2025thinkguard} distills structured critiques from larger LLMs. While these achieve strong accuracy, they incur substantial latency overhead (3--4$\times$ slower than non-reasoning baselines).

\subsection{Datasets}
\label{sec:datasets_details}
\textbf{DynaBench}~\citep{hoover2025dynaguard} consists of difficult policy-grounded safety evaluation examples. Each example includes content to moderate and a safety config with a list of policy items expressed as natural language rules. The benchmark is unusually demanding on the policy-reasoning side: each example carries on average $13.8$ policy items and up to $91$ in the heaviest configurations, so the model must read a long policy list, locate the small subset that actually applies, and emit verdict + violated index without distraction from the surrounding clauses. To probe the failure modes surfaced in Section~\ref{sec:investigation}, we report on a safety-augmented split, \textbf{DynaBench (Aug)}, that combines two complementary perturbations of the released test set:
\begin{itemize}
\item \emph{Policy shuffling.} For $50\%$ of the examples we keep the original policy list and ground-truth label but draw a uniformly random permutation of the policy items, so that the violated index moves around the list. This isolates positional bias without changing what the policy says.
\item \emph{Counterfactual unsafe$\,\to\,$safe.} For $30\%$ of the examples that are originally labeled unsafe, we strip \emph{all} policies the example violates from the policy list and flip the gold label to \texttt{safe}; the remaining (non-violated) policies are kept and the resulting list is shuffled. A model that truly grounds its verdict on the active rule set should follow the label flip, whereas a model leaning on content priors will keep predicting unsafe.
\end{itemize}
The two perturbations are sampled independently per example, so a small fraction receive both. We also keep the unaugmented split, \textbf{DynaBench (Original)}, on which we report Table~\ref{tab:results_dynabench_old}. DynaGuard was trained on the original DynaBench and is likely to overfit to its surface form: comparing its numbers across the original and augmented splits gives a direct read on how much of its headline accuracy depends on the exact policy phrasing and ordering at training time, rather than on policy-grounded reasoning.

\begin{table}[h]
\small
\caption{Results on the original DynaBench (no augmentation). \textbf{Acc}: Safety Accuracy. \textbf{F1}: Safety F1. \textbf{Latency}: average inference time in milliseconds. DynaGuard is trained on this split and is likely to overfit; the gap between this table and Table~\ref{tab:results} (DynaBench Aug) indicates how much accuracy depends on surface phrasing and ordering.}
\label{tab:results_dynabench_old}
\begin{center}
\begin{tabular}{lccc}
\toprule
\textbf{Model} & \textbf{Acc} & \textbf{F1} & \textbf{Latency (ms)} \\
\midrule
Qwen3-4B             & 56.90 & 23.02 &  232 \\
Qwen3-4B-Thinking    & 67.40 & 53.29 & 7258 \\
GuardReasoner-3B    & 58.56 & 31.19 & 4228 \\
DynaGuard-4B        & 71.82 & 69.94 & 258 \\
LPG                 & 73.90 & 61.25 & 710 \\
\bottomrule
\end{tabular}
\end{center}
\end{table}

\textbf{GuardSet-X} is a policy-grounded safety moderation benchmark where each test example is associated with exactly one policy item that determines the safety label. This one-to-one mapping enables precise evaluation of policy identification accuracy. We randomly select 2000 examples from GuardSet-X as testset.

\textbf{Out-of-distribution benchmarks.} For the OOD evaluation in Section~\ref{sec:ood}, we use three benchmarks not seen during training. \emph{HarmBench}~\citep{mazeika2024harmbench} is a red-teaming benchmark with 320 harmful behaviors spanning 7 semantic categories. \emph{WildGuardTest}~\citep{han2024wildguard} contains 1{,}725 prompt-response pairs annotated for harmfulness across 13 canonical categories; an example is unsafe iff the prompt or response is harmful. \emph{PolicyGuardBench}~\citep{wen2025towards} is a policy-grounded benchmark for detecting policy violations in web-agent trajectories, with within- and cross-subdomain policy pairings and a prefix-based detection task in addition to full-trajectory evaluation. HarmBench and WildGuardTest do not ship policy strings, so we synthesize per-example policy lists from each benchmark's native taxonomy (4--10 candidate policies per sample, always including the violated category for unsafe examples), matching the usage format of dynamic policy settings. We report Safety F1 only for these three benchmark.

\subsection{Evaluation Metrics}

We report two categories of metrics:

\textbf{Safety Metrics:} Safety Accuracy (Acc) measures binary classification correctness. Safety F1 (F1) evaluates violation detection performance, with precision measuring false positive rate and recall measuring false negative rate. Consistency Rate (CR) measures agreement across policy order permutations.

\textbf{Efficiency Metrics:} Latency reports average inference time per sample in milliseconds on an NVIDIA A100 80GB GPU with batch size 1 and temperature 0.0.

\subsection{Implementation Details}

All models are evaluated under identical conditions: NVIDIA A100 80GB GPU, batch size 1 (realistic streaming scenario), temperature 0.0 for deterministic evaluation, FP16 precision. Non-reasoning models use max tokens = 512; reasoning models use max tokens = 8192. For LPG, we use $m_1=4$ latent tokens for Stage 1 (intent) and $m_2=6$ latent tokens for Stage 2 (risk), yielding 10 total latent reasoning tokens.

\section{Detailed Training Configuration}
\label{sec:training_details}

\subsection{Model Architecture}

LPG uses Qwen3-4B as the base causal LM. We add LoRA adapters (rank $r{=}128$, $\alpha{=}32$, dropout $0.05$) to all attention and feedforward projection matrices (\texttt{q\_proj}, \texttt{k\_proj}, \texttt{v\_proj}, \texttt{o\_proj}, \texttt{up\_proj}, \texttt{down\_proj}, \texttt{gate\_proj}). Three special tokens are appended to the vocabulary: \texttt{[PAD]}, \texttt{<bot>} (begin-of-thought, appended to the question to mark the start of latent reasoning), and \texttt{<eot>} (end-of-thought, prepended to the verdict to mark the transition from latent to explicit generation).

\paragraph{Projection module.}
Between latent rollout steps, the top-layer hidden state $\mathbf{h}^{(L)}_t \in \mathbb{R}^d$ is mapped back to the embedding space via a learned MLP projector:
\begin{equation}
\text{Proj}(\mathbf{h}) = \text{LN}(\mathbf{W}_2\,\text{GELU}(\mathbf{W}_1\,\text{Dropout}(\mathbf{h}))),
\end{equation}
where $\mathbf{W}_1 \in \mathbb{R}^{d \times d_p}$, $\mathbf{W}_2 \in \mathbb{R}^{d_p \times d}$, $d_p{=}2560$, and $\text{LN}$ is LayerNorm. The same projector is shared across all latent steps and stages.

\paragraph{Explain projectors.}
For the summary reconstruction loss, each stage has a dedicated explain projector consisting of $N_{\text{proj}}$ stacked layers of Linear--GELU--LayerNorm--Dropout (each $\mathbb{R}^d \to \mathbb{R}^d$). We use $N_{\text{proj}}{=}3$ layers with dropout $0.1$. These projectors are discarded at inference time.

\subsection{Training Corpus: Composition and Curation}
\label{sec:corpus_details}

This subsection expands Section~\ref{sec:corpus} with the full per-source breakdown and the three curation principles (taxonomy normalization, per-example policy-list synthesis, and teacher-grounded reasoning generation) that turn the raw $40{,}041$-record collection into a single training distribution.

\paragraph{Per-source composition.}
The policy-grounded portion ($21{,}091$ records) supplies canonical user--policy--conversation triples: DynaBench~\citep{hoover2025dynaguard} ($13{,}500$ records) and GuardSet-X~\citep{wen2025towards} ($7{,}491$ records, structured with explicit policy organization by subdomain. The general-guardrail portion ($19{,}050$ records) draws from BeaverTails~\citep{ji2023beavertails} ($8{,}600$, 14 harm categories), Aegis-AI Content Safety v2~\citep{ghosh2025aegis} ($5{,}000$, 5-level safety taxonomy), SaladBench~\citep{li2024saladbench} ($3{,}000$ from the attack-enhanced split, 6/16/66 hierarchical taxonomy bound at the 16-task level), Toxic-Chat~\citep{lin2023toxicchat} ($2{,}000$, real-world toxicity labels), and XSTest v2~\citep{rottger2023xstest} ($450$, retained as a hard-negative over-refusal signal). The final mixture is $45.5\%$ safe / $54.5\%$ unsafe, with $7.5\%$ of unsafe examples carrying multiple violated clauses (mean $0.63$ violations per record).

\paragraph{Taxonomy normalization.}
Each source ships its own categorical taxonomy, with very different surface forms (BeaverTails' fine-grained harm labels vs.\ SaladBench's hierarchical task names vs.\ Aegis' 5-level safety taxonomy). To prevent the student from over-fitting to any specific phrasing, we rewrite every category into an imperative single-sentence rule (e.g.\ \emph{``Do not give instructions for acquiring, manufacturing, or using illegal drugs, controlled substances, or prohibited weapons''}), producing a unified ``policy book'' per source whose surface forms differ from those used in DynaBench/GuardSet-X.

\paragraph{Per-example policy-list synthesis.}
Existing guardrails latch onto policy ordering (Section~\ref{sec:investigation}). We therefore pair every training example with a freshly sampled policy list rather than reusing a fixed taxonomy. For each record we draw $K \sim \text{Uniform}\{4, 10\}$ policies from the source's policy book (\emph{always} including the violated rule(s) when unsafe), and shuffle the resulting list. With probability $0.30$ we additionally seed the list with $1$--$2$ policies drawn from \emph{other} sources' books, so that the model must reason about heterogeneous policy phrasings within a single example (cross-taxonomy mixing). Stratified subsampling along (safe/unsafe $\times$ category) keeps long-tail categories from being starved during training. The combined effect is that the model effectively never sees the same policy list twice across the $40$k corpus, removing the positional shortcut exposed in Section~\ref{sec:investigation} from the training distribution by construction.

\paragraph{Teacher-grounded reasoning generation.}
For every record, an offline pass through a Qwen3-32B teacher~\citep{yang2025qwen3} produces (i)~the explicit structured trace $\langle$\texttt{Intent}, \texttt{Risk}, \texttt{Output}$\rangle$ used to supervise the reference reasoning loss $\mathcal{L}_{\text{ref}}$ and the answer-position distillation, and (ii)~the per-stage \texttt{IntentSummary} / \texttt{RiskSummary} targets used by the explain loss $\mathcal{L}_{\text{explain}}$. The teacher is conditioned on the ground-truth verdict $(y, P^*)$ so that its rationale is constrained to \emph{justify the correct decision} rather than reason free-form; this eliminates reasoning--label drift that would otherwise propagate into the student's latents. Summaries are capped at $192$ tokens to enforce compression.

\subsection{Training Data Pipeline}

\paragraph{Teacher reasoning format.}
The Qwen3-32B teacher generates explicit structured reasoning for each training example in the format: \texttt{<Intent>...</Intent>$\backslash$n$\backslash$n<Risk>...</Risk>$\backslash$n$\backslash$n<Output>safe / unsafe, policy...</Output>}. The Intent block analyzes the user's true intent including hidden jailbreak attempts, and the Risk block identifies relevant policy items and provides a concise violation rationale.

\paragraph{Summary cache generation.}
A separate offline step uses the same Qwen3-32B teacher to compress each explicit reasoning stage into a compact summary target. The teacher is prompted to produce two tagged blocks: \texttt{<IntentSummary>} (user intent and any hidden malicious goal) and \texttt{<RiskSummary>} (minimum policy-relevant evidence and concise rationale). Summaries are capped at 192 tokens and cached in augmented JSONL files to avoid repeated teacher inference during training.

\paragraph{Data preprocessing.}
For each training example, three parallel token sequences are constructed:
\begin{enumerate}
\item \textbf{Encoder path} (for latent rollout): the question (\texttt{annotation\_input}) followed by \texttt{<bot>}.
\item \textbf{Decoder path} (for explicit verdict output): \texttt{<eot>} followed by the \texttt{<Output>} block with the JSON verdict. Standard next-token cross-entropy labels are applied.
\item \textbf{Reference path} (for teacher supervision): the full sequence of question + explicit reasoning + output. Labels mask the question prefix (set to $-100$) so that only the reasoning and verdict tokens contribute to $\mathcal{L}_{\text{ref}}$.
\end{enumerate}
Additionally, \emph{stage boundary positions} (the token indices of \texttt{</Intent>} and \texttt{</Risk>} in the reference sequence) are precomputed for the stage-boundary distillation sub-loss. Examples exceeding 800 tokens (question + reasoning combined) are filtered out.

\subsection{Training Hyperparameters}

Training uses DeepSpeed ZeRO-2 across 4 NVIDIA A100-80G GPUs. Table~\ref{tab:hyperparams} summarizes the full configuration.

\begin{table}[h]
\small
\caption{LPG training hyperparameters.}
\label{tab:hyperparams}
\begin{center}
\begin{tabular}{ll}
\toprule
\textbf{Hyperparameter} & \textbf{Value} \\
\midrule
Base model & Qwen3-4B \\
LoRA rank / alpha & 128 / 32 \\
LoRA target modules & All attention + FFN projections \\
Latent tokens per stage $(m_1, m_2)$ & 4 (intent), 6 (risk) \\
Projection hidden dim $d_p$ & 2560 \\
Explain projector layers & 3 \\
\midrule
Learning rate & $1 \times 10^{-6}$ \\
LR scheduler & Linear \\
Warmup ratio & 0.10 \\
Weight decay & 0.1 \\
Max gradient norm & 2.0 \\
Precision & BF16 \\
Optimizer & AdamW \\
\midrule
Epochs & 3 \\
Per-device batch size & 2 \\
Gradient accumulation steps & 8 \\
Effective batch size & 64 (4 devices $\times$ 2 $\times$ 8) \\
\midrule
$\lambda_{\text{out}}$ (verdict CE) & 2.0 \\
$\lambda_{\text{distill}}$ (teacher hidden-state distillation) & 10.0 \\
$\beta$ (stage-boundary sub-weight inside $\mathcal{L}_{\text{distill}}$) & 0.1 \\
$\lambda_{\text{ref}}$ (reference reasoning CE) & 1.0 \\
$\lambda_{\text{explain}}$ (summary reconstruction) & 0.5 \\
Distillation loss function & Smooth L1 \\
Max token length (filter threshold) & 800 \\
Summary max target length & 192 \\
\bottomrule
\end{tabular}
\end{center}
\end{table}

\subsection{Training Algorithm}
\label{sec:training_algorithm}

Algorithm~\ref{alg:lpg_train} provides the full pseudocode for the multi-objective LPG training procedure summarized in Section~\ref{sec:training}. The algorithm interleaves the student's latent reasoning rollout with a teacher reasoning pass and computes the four loss terms (output, teacher hidden-state distillation, reference, and explain) in a single phase, where the distillation term itself combines the answer-position and stage-boundary sub-losses.

\begin{algorithm}[h]
\caption{Training LPG latent reasoning with multi-objective teacher supervision}
\label{alg:lpg_train}
\begin{algorithmic}[1]
\REQUIRE Dataset $\mathcal{D}$ of contexts $\mathbf{c}=(\mathcal{P},x)$; teacher $f_{\theta_T}$; student $f_\theta$ with LoRA adapters; explain projectors $\{\text{Proj}_k\}_{k\in\{1,2\}}$; projection module $\text{Proj}$; latent lengths $(m_1,m_2)$; loss weights $(\lambda_{\text{out}}, \lambda_{\text{distill}}, \lambda_{\text{ref}}, \lambda_{\text{explain}})$ and stage-boundary sub-weight $\beta$.
\STATE Initialize $f_\theta$ from a base LLM with LoRA adapters; initialize projector parameters.
\FOR{each minibatch $\mathbf{c} \sim \mathcal{D}$}
\STATE \textbf{// Latent reasoning path (student)}
\STATE Encode question $\mathbf{c}$ with $f_\theta$; extract final hidden state $\mathbf{h}_0 = \mathbf{h}^{(L)}_{|\mathbf{c}|}$.
\STATE $\mathbf{e}_1 \leftarrow \text{Proj}(\mathbf{h}_0)$ \COMMENT{Project to embedding space}
\FOR{each stage $k \in \{1, 2\}$}
    \FOR{$j = 1, \ldots, m_k$}
        \STATE $\mathbf{h}_j^{(k)} \leftarrow f_\theta(\mathbf{e}_j; \text{KV cache})$; \quad $\mathbf{e}_{j+1} \leftarrow \text{Proj}(\mathbf{h}_j^{(k)})$
    \ENDFOR
    \STATE Store $\mathbf{z}^{(k)}_{m_k} \leftarrow \mathbf{h}_{m_k}^{(k)}$ \COMMENT{Final hidden for stage-boundary distillation}
    \STATE Compute $\mathcal{L}_{\text{explain}}^{(k)}$: project $\mathbf{Z}^{(k)}$ via $\text{Proj}_k$, reconstruct summary $\mathbf{s}^{(k)}$ through $f_\theta$
\ENDFOR
\STATE Generate verdict tokens from KV cache; compute $\mathcal{L}_{\text{out}}$ (Eq.~\ref{eq:out_loss}).
\STATE \textbf{// Teacher reasoning path}
\STATE Run $f_\theta$ on full explicit reasoning sequence (no grad); extract boundary states $\mathbf{h}^{(L)}_{\text{teacher}}[b_k]$.
\STATE Run $f_\theta$ on full explicit reasoning sequence (with grad); compute $\mathcal{L}_{\text{ref}}$ (Eq.~\ref{eq:ref_ce}).
\STATE \textbf{// Teacher hidden-state distillation}
\STATE Compute $\mathcal{L}_{\text{distill}}^{\text{ans}}$ at answer positions across all layers and $\mathcal{L}_{\text{distill}}^{\text{stage}}$ at stage boundaries (Eq.~\ref{eq:distill}).
\STATE $\mathcal{L}_{\text{distill}} \leftarrow \mathcal{L}_{\text{distill}}^{\text{ans}} + \beta\,\mathcal{L}_{\text{distill}}^{\text{stage}}$.
\STATE Update $\theta$ using gradients of $\mathcal{L}$ (Eq.~\ref{eq:total_loss}).
\ENDFOR
\end{algorithmic}
\end{algorithm}

\subsection{Inference Procedure}

At inference time, the model performs the following steps:
\begin{enumerate}
\item \textbf{Prompt encoding}: The question is tokenized and appended with \texttt{<bot>}. A forward pass through the base LM produces the final hidden state at the prompt boundary.
\item \textbf{Latent rollout}: The hidden state is projected via the MLP projector and fed back as input for $m_1{=}4$ latent steps (intent stage), followed by $m_2{=}6$ latent steps (risk stage). Each step uses KV-cache for efficient incremental computation. No discrete tokens are generated during this phase.
\item \textbf{Verdict generation}: The \texttt{<eot>} token embedding is appended, and the model autoregressively emits the compact verdict string, one of \texttt{``safe''}, \texttt{``unsafe, policy $n$''}, or \texttt{``unsafe, policy $n_1, n_2, \ldots$''}.
\item \textbf{Verdict extraction}: A deterministic regex parser maps the compact string to $(y, P^*)$. (Note: although the teacher reasoning collected at training time uses a JSON \texttt{<Output>} block, the student is trained on--and emits--the compact form, which is shorter and equally parseable.)
\end{enumerate}
Special tokens (\texttt{[PAD]}, \texttt{<bot>}, \texttt{<eot>}) are suppressed during verdict generation by setting their logits to $-\infty$. The explain projectors and reference reasoning path are not used during inference.


\section{Decoding-Variance Sweep}
\label{sec:variance}

To check that the statistically significance of results in Table~\ref{tab:results} are not artefacts of a single sampled trajectory, we run a small decoding-variance study. For each (model, dataset) cell we sweep three temperatures $T \in \{0.3, 0.7, 1.0\}$ and draw five independent decoding seeds at each temperature, for a total of $n{=}15$ runs per cell. Sampling is forced on for every model so the stochasticity is observable and comparable, and the random seeds are reset before each rerun. The evaluation set is a subset of $200$ examples: $100$ uniformly sampled from GuardSet-X and $100$ from the DynaBench (Aug) split.

Since each run reduces to $100$ correct/incorrect outcomes per dataset, we keep the report simple and use Safety Accuracy as the single metric. We report the across-run standard deviation $\sigma$ and the half-width of the $95\%$ confidence interval ($1.96 \cdot \sigma / \sqrt{n}$ with $n{=}15$), in percentage points.

\begin{table}[h]
\small
\setlength{\tabcolsep}{6pt}
\caption{Decoding-variance sweep on a $100{+}100$ subset of GuardSet-X and DynaBench (Aug). Each cell aggregates $n{=}15$ runs ($3$ temperatures $\times$ $5$ reruns). We report the across-run standard deviation $\sigma$ and the $95\%$ CI half-width for Safety Accuracy (in percentage points).}
\label{tab:variance}
\begin{center}
\begin{tabular}{lcc|cc}
\toprule
Model & \multicolumn{2}{c|}{GuardSet-X (100)} & \multicolumn{2}{c}{DynaBench Aug (100)} \\
\cmidrule(lr){2-3} \cmidrule(lr){4-5}
 & Acc $\sigma$ & Acc CI$_{95}$ & Acc $\sigma$ & Acc CI$_{95}$ \\
\midrule
Qwen3-4B          & 0.00 & 0.00 & 0.05 & 0.03 \\
Qwen3-4B-Thinking & 2.18 & 1.10 & 2.65 & 1.34 \\
DynaGuard-4B      & 1.02 & 0.52 & 0.71 & 0.29 \\
GuardReasoner-3B  & 2.27 & 1.15 & 1.83 & 0.93 \\
LPG-4B (Ours)     & 0.51 & 0.26 & 0.27 & 0.14 \\
\bottomrule
\end{tabular}
\end{center}
\end{table}

Two observations follow. First, the variance pattern tracks how much each model commits to the decoder. Qwen3-4B emits a near-deterministic short verdict, so different temperatures and seeds collapse to the same answer and $\sigma$ is essentially zero on both datasets. The two explicit-reasoning baselines, Qwen3-4B-Thinking and GuardReasoner-3B, sample long chains of reasoning tokens before committing to a verdict, and any token-level divergence may flip the eventual label, so they record the largest CIs (Qwen3-4B-Thinking is the noisiest on DynaBench Aug at $1.34$ and GuardReasoner-3B is the noisiest on GuardSet-X at $1.15$). DynaGuard-4B sits in the middle: its verdict is short, but its hardcoded internal sampling leaves $\sim$$0.5$--$1$ point of residual jitter. LPG-4B is close to the deterministic floor: even with sampling forced on, the policy-anchored compact verdict gives the decoder very few degrees of freedom, and the latent reasoning stage produces no sampled tokens at all. Second, the across-run intervals are far smaller than the cross-model gaps in Table~\ref{tab:results}: among the five models in this sweep, LPG's Accuracy lead is $15.85$ points over the next-best model on GuardSet-X and $7.18$--$13.45$ points on DynaBench (Aug), while every CI$_{95}$ half-width above is at most $1.34$. The headline ranking is therefore stable under decoding noise, and the variance budget is well below the magnitude of the reported gains.


\section{Limitations and Broader Impacts}
\label{sec:limitations}

LPG already establishes a strong accuracy-latency Pareto front for policy-grounded safety moderation, and the design choices that make this possible double as natural directions for further work. The latent-budget sweep continues to rise monotonically beyond the deployed ten tokens, so latency-tolerant settings such as offline batch moderation can be served from a single checkpoint trained with a curriculum over budgets. Because the latent training pipeline is backbone-agnostic and the latent budget is independent of backbone size, scaling LPG to larger backbones should compound the accuracy advantage without changing the latency profile, and stronger teachers (larger reasoning models or richer corpora) plug in without architectural change.

Looking beyond safety moderation, the latent compression architecture introduced here is an early instance of a broader \emph{latent compliance reasoning} paradigm: any deployed AI system that must condition on natural-language rules supplied at inference time, ranging from agentic tool-use guards and jurisdiction-aware regulatory compliance to enterprise content-policy enforcement and constitutional-AI runtime alignment, faces the same accuracy-latency tension that motivated LPG. We see two especially attractive follow-ups: integrating LPG-style latent reasoning into agent frameworks as a low-latency action-time guard over tool calls, and co-evolving policy and student so that clause-attribution feedback collected from deployment refines the policy over time.

\paragraph{Broader impacts.}
\label{sec:broader_impacts}
LPG is designed for policy-grounded content moderation, where the safety policy is supplied at inference time. The most direct positive impact is that latency-constrained deployments such as real-time chat moderation, on-device safety filters, and high-volume customer-facing applications can now afford the kind of policy-aware reasoning that was previously restricted to slow explicit-reasoning systems. The clause-anchored verdict format also improves transparency: each unsafe decision points to specific policy items, enabling operators and end users to audit, contest, and refine the deployed policy without retraining the model. 

As with any safety moderation system, failure modes warrant attention. For instance, false positives can over-restrict legitimate communication; LPG mitigates this by emitting clause-level attributions for every unsafe verdict, so operators can identify and refine the offending clause rather than having to retrain the model. We view the combination of fast policy reasoning, clause-level attribution, and runtime policy injection as net positive for the safety-and-transparency stack of deployed AI systems.

\end{document}